\documentstyle[twocolumn,pre,aps,psfig]{revtex}
\begin{document}
\draft

\title{Controlling Physical Systems with Symmetries}
\author{R. O. Grigoriev and M. C. Cross}
\address{Condensed Matter Physics 114-36,\\
California Institute of Technology, Pasadena CA 91125 }
\date{\today}
\maketitle

\begin{abstract}
 Symmetry properties of the evolution equation and the state to be
controlled are shown to determine the basic features of the linear control
of unstable orbits. In particular, the selection of control parameters and
their minimal number are determined by the irreducible representations of
the symmetry group of the linearization about the orbit to be controlled. 
We use the general results to demonstrate the effect of symmetry on the
control of two sample physical systems: a coupled map lattice and a
particle in a symmetric potential.
 \end{abstract}


\narrowtext

\section{Introduction}

Despite the recent wave of interest towards controlling chaotic dynamics
\cite{romeiras,petrov} an interesting and important question of
controlling systems with symmetries received surprisingly little attention
in the physics literature. The importance of symmetries in controlling,
for instance, spatiotemporal chaos is evident, since the systems typically
show rotational and translational symmetries.  Although the presence of
symmetries usually significantly simplifies the analysis of system
dynamics, it also makes control schemes more complicated due to the
inherent degeneracies of the evolution operators. In fact, the presence of
symmetries, explicit or implicit, makes a number of
single-control-parameter methods fail \cite{romeiras,ding}, calling for
multi-parameter control \cite{barreto,warncke,self,locher}. 

In order to see how these restrictions arise, let us consider a general
discrete-time system (the arguments for continuous-time systems are very
similar), whose evolution is described by the map $F: R^N\times
R^M\rightarrow R^N$,
 \begin{equation}
 \label{eq_ev_nl}
 {\bf z}^{t+1}=F({\bf z}^t,{\bf p}),
 \end{equation}
 where ${\bf z}$ is an $N$-dimensional state vector and ${\bf p}$ is an
$M$-dimensional parameter vector. Linearizing about the steady state
solution of ${\bf z}^*=F({\bf z}^*,{\bf p}^*)$ and denoting ${\bf
x}^t={\bf z}^t-{\bf z}^*$ and ${\bf u}^t={\bf p}^t-{\bf p}^*$, we readily
obtain
 \begin{equation}
 \label{eq_ev_lin}
 {\bf x}^{t+1}=A{\bf x}^t+B{\bf u}^t,
 \end{equation}
 where $A_{ij}=\partial F_i({\bf z}^*,{\bf p}^*)/\partial z_j$ is the
Jacobian of the transformation and $B_{ij}=\partial F_i({\bf z}^*,{\bf
p}^*)/\partial p_j$ is the control matrix.

If the steady state ${\bf z}^*$ is unstable, it can be stabilized by an
appropriate feedback through the time-dependent control perturbation ${\bf
u}^t$ if the matrices $A$ and $B$ satisfy certain conditions. We will
understand the design of the control scheme as an appropriate choice of
the control matrix $B$. We will see below that the conditions affecting
this choice can be easily obtained from the symmetry properties of the
system and the controlled state. 

\section{Stabilizable vs. Controllable Systems}

Assuming that the feedback is linear in the deviation from the steady
state ${\bf z}^*$ we can write
 \begin{equation}
 \label{eq_feedback}
 {\bf u}^t=-K{\bf x}^t,
 \end{equation}
and one obtains the linearized evolution equation in the following form:
 \begin{equation}
 \label{eq_lin_contr}
 {\bf x}^{t+1}=(A-BK){\bf x}^t.
 \end{equation}
 The matrix $A'=A-BK$ has stability properties different from the
stability properties of the matrix $A$. This can be exploited to make the
steady state ${\bf z}^*$ (the matrix $A-BK$) stable under control. 

The dynamical system (\ref{eq_ev_lin}) or the pair $(A,B)$ is said to be
{\sl stabilizable}, if there exists a state feedback (\ref{eq_feedback}),
such that the system (\ref{eq_lin_contr}) is stable. The stabilizability
is the property, which often sensitively depends on the values of the
control parameters ${\bf p}^*$. 

In the majority of practical situations it is preferable to have an
adaptive control that would stabilize a given steady state ${\bf z}^*({\bf
p}^*)$ for arbitrary values of the system parameters. This is especially
important if one is to track the trajectory ${\bf z}^*$ as ${\bf p}^*$
changes or use the same control arrangement to stabilize different steady
(or even periodic) states.

Such a control scheme is obtained if the more restrictive condition of
{\sl controllability} is imposed on the matrices $A$ and $B$. The
dynamical system (\ref{eq_ev_lin}) or the pair $(A,B)$ is said to be {\sl
controllable} if the eigenvalues of the matrix $A-BK$ can be freely
assigned (with complex ones in conjugate pairs), which is equivalent (see
Theorem 5.13 in Reference \cite{rubio}) to requiring that: 
 \begin{equation}
 \label{eq_contr}
 {\rm rank}(C)=N,
 \end{equation}
 where $C=(B\ AB\ A^2B\ \cdots\ A^{N-1}B)$ is called the {\sl
controllability matrix}. Relation (\ref{eq_contr}) was introduced into the
physics literature from linear systems theory by Romeiras {\sl et al.}
\cite{romeiras} as a simple, but practical test of the controllability. 

In order to better understand the restrictions imposed by the symmetry, it
is beneficial to look at the controllability condition written in the form
(\ref{eq_contr}) from the geometrical point of view assuming $M=1$ and
$B={\bf b}$. Suppose we let the system evolve under control for $\tau$
steps from the initial state ${\bf x}^t$. The final state will be given by
 \begin{equation}
 {\bf x}^{t+\tau}=A^\tau{\bf x}^t+\sum_{m=0}^{\tau-1}
  A^{\tau-1-m}{\bf b}\,u^{t+m}.
 \end{equation}
 The controllability in this context is equivalent to the vectors
 \begin{equation}
 \label{eq_bas_set}
 {\bf f}^m=A^{\tau-1-m}{\bf b},\quad m=0,\cdots,\tau-1
 \end{equation}
 spanning the state space for $\tau=N$, so that any initial state can
be mapped to any final state in $\tau$ time steps by an appropriate choice
of the ``coordinates'' $u^{t+m}$ of the difference ${\bf
x}^{t+\tau}-A^\tau{\bf x}^t$. 

If the matrix $A$ is non-degenerate (has a non-degenerate spectrum), one
can always find a vector ${\bf b}$, such that the resulting set
(\ref{eq_bas_set}) forms a basis. However, if $A$ is degenerate (which is
a usual consequence of symmetry), there will exist an invariant subspace
$L^r\subset R^N$, with the dimension ${\rm dim}(L^r)>1$, where the
dynamics of the system cannot be controlled with just one control
parameter (see Reference \cite{romeiras} for an example of such a
situation).

If the system dynamics in $L^r$ happens to be stable, the system can still
be stabilized similarly to the non-degenerate case, but we have to ensure
controllability in case the dynamics in this subspace is unstable.  This
can be achieved by increasing the number of control parameters $M$, which
extends the set (\ref{eq_bas_set}), until it spans every unstable
invariant subspace of $R^N$. This would lead one to assume that $M$ should
be defined by the dimension of the largest invariant subspace, or
equivalently the highest degeneracy of the Jacobian matrix $A$. We will
see however, that various kinds of degeneracy have a somewhat different
effect on the controllability of the system.

\section{The Number and Selection of Parameters}

Let us assume that the evolution equation (\ref{eq_ev_nl}) possesses a
symmetry described by a symmetry group $\mathcal{G}$, i.e. the map $F$
commutes with all group actions: 
 \begin{equation}
 \label{eq_equiv}
 F(g({\bf z}),{\bf p})=g(F({\bf z},{\bf p})),
 \quad \forall g\in \mathcal{G},
 \end{equation}
 or in other words, the function $F({\bf z},{\bf p})$ is
$\mathcal{G}$-equivariant.

The symmetry of the linearized equation (\ref{eq_ev_lin}) in the absence
of control (${\bf u}=0$) is in general different from (although closely
related to) the symmetry of the full nonlinear equation (\ref{eq_ev_nl}).
We will call the respective symmetry group $\mathcal{G}^*$. It generates
the matrix representation $T$ in the state space $R^N$:
 \begin{equation}
 \label{eq_in_rep}
 (g({\bf x}))_i=(T(g){\bf x})_i=T(g)_{ij}x_j.
 \end{equation}
Due to the symmetry all matrices $T(g)$ commute with the Jacobian
 \begin{equation}
 \label{eq_commut}
 T(g)A=AT(g),\quad \forall g\in \mathcal{G}^*.
 \end{equation}

$\mathcal{G}^*$ depends on both $\mathcal{G}$ and the reference state
${\bf z}^*$, or to be exact, its symmetry group, which we denote
$\mathcal{H}$: 
 \begin{equation}
 h({\bf z}^*)={\bf z}^*,\quad\forall h\in \mathcal{H}.
 \end{equation}
 The symmetry of the evolution equation is reduced upon linearization, if
the reference state has low symmetry, and then $\mathcal{G}^*$ becomes one
of the subgroups of $\mathcal{G}$. On the other hand, $\mathcal{G}^*$
might be equal to $\mathcal{G}$ or even include $\mathcal{G}$ as a
subgroup for highly symmetric reference states, with the apparent symmetry
increased by linearization.

Decomposing $T$ into a sum of irreducible representations $T^r$ of the
group $\mathcal{G}^*$ with respective dimensionalities $m_r$,
$r=1,\cdots,l$ we obtain:
 \begin{equation}
 T=\sum_{r\oplus} T^r, \qquad N=\sum_r m_r.
 \end{equation}
 According to the standard group theoretic analysis \cite{hammer} the
Jacobian $A$ will have eigenvalues $\lambda_r$ with multiplicities $n_r\ge
m_r$, corresponding to the dimensions of the irreducible representations
$T^r$ contained in the decomposition of $T$. If $m_r>1$ for some $r$, the
Jacobian becomes degenerate, which causes certain control methods to fail
(see, for example, Reference \cite{ding}). 

Next we use the result of linear systems theory based on the Jordan
decomposition of the Jacobian matrix:
 \begin{equation}
 \Lambda=SAS^{-1}=\left(\matrix{
    \Lambda^1 & & & \cr
    & \Lambda^2 & & \cr
    & & \ddots & \cr
    & & &\Lambda^l }\right),
 \end{equation}
 where the Jordan superblock 
 \begin{equation}
 \Lambda^r=\left(\matrix{
    \Lambda^{r1} & & & \cr
    & \Lambda^{r2} & & \cr
    & & \ddots & \cr   
    & & &\Lambda^{rs_r} }\right)
 \end{equation}
 corresponding to the eigenvalue $\lambda_r$ has dimension $n_r$ and
consists of $s_r$ Jordan blocks
 \begin{equation}
 \Lambda^{ri}=\left(\matrix{
    \lambda_r & & & & \cr
    1 & \lambda_r & & & \cr
    & \ddots & \ddots & & \cr   
    & & 1 & \lambda_r & \cr
    & & & 1 & \lambda_r }\right).
 \end{equation}

Since the controllability is invariant with respect to coordinate
transformations (Theorem 5.17 in Reference \cite{rubio}), condition
(\ref{eq_contr}) is satisfied for the pair $(A,B)$ if and only if it is
satisfied for the pair $(\Lambda,\hat{B})$, where $\hat{B}=SB$ is the
transformed control matrix. 
 
If $\hat{B}$ is partitioned according to the block structure of $\Lambda$: 
 \begin{equation}
 \hat{B}=\left(\matrix{
 \hat{B}^1\cr\hat{B}^2\cr\vdots\cr\hat{B}^l}\right),\quad
 \hat{B}^r=\left(\matrix{
 \hat{B}^{r1}\cr\hat{B}^{r2}\cr\vdots\cr\hat{B}^{rs_r}}\right),\quad
 \hat{B}^{ri}=\left(\matrix{
 \hat{b}^{ri}_1\cr\hat{b}^{ri}_2\cr\vdots\cr\hat{b}^{ri}_{n_{ri}}}\right)
 \end{equation}
 the controllability condition for the pair $(\Lambda,\hat{B})$ is reduced
to the controllability conditions for each pair $(\Lambda^r,\hat{B}^r)$,
which, in turn, is satisfied (Theorem 5.18 in Reference \cite{rubio}) if
and only if for each $r$ the set of $s_r$ $M$-dimensional row vectors
 \begin{equation}
 \label{eq_row_vec}
 \hat{b}^{r1}_1,\hat{b}^{r2}_1,\cdots,\hat{b}^{rs_r}_1
 \end{equation}
 is linearly independent. This in turn can be achieved if and only if
$M\ge s_r$ for every $r$. Hence the minimal number of control parameters
should equal the maximal number of Jordan blocks contained in any one
superblock: 
 \begin{equation}
 \label{eq_max_blk}
 M_{min}=\max_r s_r.
 \end{equation}

In general, the system under consideration is not Hamiltonian and
therefore its Jacobian matrix $A$ is not hermitian, hence
non-diagonalizable. Therefore we have $n_r\ge s_r\ge m_r$. However in the
absence of accidental degeneracies $n_r=s_r=m_r$ for every $r$, and the
condition (\ref{eq_row_vec}) is equivalent to
 \begin{equation}
 \label{eq_rank}
 {\rm rank}(\hat{B}_r)=m_r.
 \end{equation}

The calculation of the transformation $S$ can be avoided for compact
groups $\mathcal{G}^*$ by using the projection operator $P^r$ on the
subspace $L^r\subset R^N$, that transforms according to the $r$-th
irreducible representation: 
 \begin{equation}
 P^r=m_r\int_{\mathcal{G}^*}\chi^r(g) T(g)\,d\mu(g).
 \end{equation}
 Here $\chi^r(g)$ is the character of the group element $g$ in the
representation $T^r$ and $d\mu(g)$ is the group measure \cite{hammer}. For
finite groups this integral is replaced with the sum.

Using the fact that
 \begin{equation}
 {\rm rank}(\tilde{B}^r)={\rm rank}(\hat{B}^r),
 \end{equation}
 where $\hat{B}^r=P^rB$, we conclude (assuming there are no accidental
degeneracies), that the controllability condition is satisfied whenever
 \begin{equation}
 \label{eq_min_num}
 M\ge \max_r m_r
 \end{equation}
 and $m_r$ of $M$ columns of $\tilde{B}^r$ are linearly independent, i.e. 
form a basis in the eigenspace $L^r$ (columns of $\tilde{B}^r$ span $L^r$)
for every $r$.

Therefore the minimal number of independent control parameters $M_{min}$
is equal to the dimensionality $m_r$ of the largest irreducible
representation $T^r$ present in the decomposition of the representation
$T$ of the group $\mathcal{G}^*$ in the state space $R^N$.

Regarding the control matrix $B$ as a row of $M$ vectors
 \begin{equation}
 B=({\bf b}_1\ {\bf b}_2\ \cdots\ {\bf b}_M),
 \end{equation}
 we see that the control scheme yielding a controllable system is obtained
by choosing the vectors ${\bf b}_i$, such that $m_r$ of the projections
$P^r {\bf b}_i$ would be linearly independent for every $r$. 

If accidental degeneracies are present, symmetry properties only give a
lower bound on the number of required control parameters and one should
look at the Jordan block structure of the Jacobian to determine the
controllability using the more general conditions (\ref{eq_max_blk}) and
(\ref{eq_row_vec}).

It is easy to see intuitively why the number of control parameters is
determined by the number of the Jordan blocks $s_r$ and not the
multiplicity $n_r$, if we look at the Jacobian already reduced to the
Jordan form. For instance,
 \begin{equation}
 A_1=\left(\matrix{
    \lambda & & \cr
    & \lambda & \cr
    & &\lambda }\right)
 \end{equation}
 generates the set of three linearly dependent vectors ${\bf f}^0={\bf
b},{\bf f}^1=\lambda{\bf b},{\bf f}^2=\lambda^2{\bf b}$ (compare to
(\ref{eq_bas_set})), that span the 1-dimensional subspace of $R^3$ for any
choice of ${\bf b}$. As a result, 3 linearly independent vectors ${\bf
b}_1,{\bf b}_2,{\bf b}_3$ are necessary to control the system.

On the contrary, the Jacobian
 \begin{equation}
 A_2=\left(\matrix{
    \lambda & & \cr
    1 & \lambda & \cr
    & 1 &\lambda }\right)
 \end{equation}
generates the linearly independent set of basis vectors that spans $R^3$,
requiring just one control vector ${\bf b}$.

Finally we should note, that symmetry does not always make the Jacobian
degenerate, and the non-degenerate case can be handled in the same way as
the one with no symmetries. Neither does the degeneracy by itself imply,
that multi-parameter control is required. Even if $n_r>0$ for some $r=r'$
(there is a degeneracy), but $s_r=m_r=1$ for every $r$ (the degeneracy is
accidental and the Jordan block $\Lambda_{r'}$ is not
block-diagonalizable), one control parameter is indeed sufficient to make
the system controllable.

\section{Continuous-Time Systems and Periodic Orbits}

The obtained results hold for continuous-time systems and can be easily
generalized to periodic trajectories. We should first observe that a
periodic trajectory of period $\tau$ can be treated as a fixed point
solution of the superposition of $\tau$ maps. The equation ${\bf z}^*=
F^\tau({\bf z}^*,{\bf p}^*)$ has $\tau$ solutions corresponding to the
points of the periodic orbit ${\bf z}^*_k$, $k=1,\cdots,\tau$. 

Next, we define the time-dependent single-step Jacobian
 \begin{equation}
 A^k_{ij}=\frac{\partial F_i({\bf z}_k^*,{\bf p}^*)}{\partial z_j}
 \end{equation}
 and the control matrix
 \begin{equation}
 B^k_{ij}=\frac{\partial F_i({\bf z}_k^*,{\bf p}^*)}{\partial p_j}.
 \end{equation}
 The controllability condition for the periodic orbit can be generalized
requiring that the pairs $(A_k,B_k)$ be controllable for every $k$. 
 
Now, suppose that the symmetry of the state ${\bf z}^*_k$ is described by
the group ${\mathcal H}_k$, such that ${\mathcal H}_k\subseteq
\mathcal{G}$. We can then write
 \begin{equation}
 h({\bf z}^*_{k+1})=h(F({\bf z}^*_k))=F(h({\bf z}^*_k))=F({\bf z}^*_k)=
 {\bf z}^*_{k+1}
 \end{equation}
 for every $h\in {\mathcal H}_k$, or consequently
 \begin{equation}
 {\mathcal H}_1\subseteq {\mathcal H}_2\subseteq\cdots\subseteq
 {\mathcal H}_\tau\subseteq {\mathcal H}_1,
 \end{equation}  
 i.e. the symmetry group of all the states of the trajectory is the
same and can be determined using ${\bf z}^*_k$ with an arbitrary $k$:
${\mathcal H}={\mathcal H}_k$.

This, in turn, means that $\mathcal{G}^*$ too is unique for any given
periodic trajectory as is the representation $T$. It is therefore enough
to know the symmetry properties of an arbitrary point of the periodic
trajectory in order to establish the requirements on the control scheme
identically to the time-independent case.

And finally, consider a continuous-time evolution equation. It can
generally be written as
 \begin{equation}
 \label{eq_ev_nl_cont}
 \partial_t{\bf z}(t)=F({\bf z}(t),{\bf p}),
 \end{equation}

Linearization of (\ref{eq_ev_nl_cont}) around the steady state ${\bf
z}^*$, similarly to the discrete-time case, yields
 \begin{equation}
 \label{eq_ev_lin_cont}
 \partial_t{\bf x}(t)=A{\bf x}(t)+B{\bf u}(t)
 \end{equation}
 and we again denote $\mathcal{G}^*$ as the group of all actions commuting
with the action of the Jacobian. The controllability of the pair $(A,B)$
ensures that all eigenvalues of $A-BK$ can be chosen negative, so that the
steady state becomes stable. As a result, the control matrix $B$ should
satisfy the same conditions as those obtained for the discrete-time case. 

\section{Coupled Map Lattice}

Next we apply the general results to the case of the coupled map lattice
defined by the following evolution equation
 \begin{equation}
 \label{eq_cml}
 z_i^{t+1}=\epsilon f(z_{i-1}^t)+(1-2\epsilon)f(z_i^t)
 +\epsilon f(z_{i+1}^t), 
 \end{equation}
 with $i=1,2,\cdots,N$ and the periodic boundary conditions, i.e. 
$z_{i+N}^t=z_i^t$ imposed. The local map function $f(x)$ can be chosen
arbitrarily.

The symmetry group $\mathcal{G}$ of the lattice includes translations by
an integer number of lattice sites (periodic boundary conditions make the
group finite) and reflections about any site. The corresponding point
group is $C_{Nv}$. It has five irreducible representations
$T^1,\cdots,T^5$ with $m_1=m_2=m_3=m_4=1$ and $m_5=2$. 

Linearizing eq. (\ref{eq_cml}) about the steady state ${\bf z}^*$ we
obtain eq. (\ref{eq_ev_lin}) with $A=CD$, where
 \begin{equation}
 C_{ij}=(1-2\epsilon)\delta_{i,j}+\epsilon(\delta_{i,j-1}+\delta_{i,j+1}) 
 \end{equation}
 (with $\delta_{i,j\pm 1}$ extended to comply with periodic b.c.) and
 \begin{equation}
 D_{ij}=f'(z^*_i)\delta_{i,j}. 
 \end{equation}
 This partition of the Jacobian into the product of two matrices
explicitly shows how the symmetry group $\mathcal{G}^*$ depends on the
symmetries of the nonlinear evolution equation and the controlled state
${\bf z}^*$. The matrix $C$ has all the symmetries imposed by the chosen
inter-site couplings of the nonlinear model: 
 \begin{equation}
 T(g)C=CT(g),\quad \forall g\in \mathcal{G},
 \end{equation}
 while the matrix $D$ reflects the symmetries of the reference state ${\bf
z}^*$: 
 \begin{equation}
 T(h)D=DT(h),\quad \forall h\in \mathcal{H},  
 \end{equation}
 and since the Jacobian $A$ only commutes with matrices that commute with
both $C$ and $D$, $\mathcal{G}^*$ should be a maximal subgroup of
$\mathcal{G}$ and $\mathcal{H}$. 

Deriving the restrictions on the control matrix $B$ is the next step. 
Once the group $\mathcal{G}^*$ is determined, we construct its
$N$-dimensional representation $T$ and decompose it into the sum of the
irreducible representations of $C_{Nv}$. For instance, a zig-zag state
gives ${\mathcal G}^*=C_{nv}$, $n=N/2$ and $M=m_5=2$; a space-period-$s$
not reflection-invariant state corresponds to ${\mathcal G}^*=C_n$,
$n=N/s$ and $M=m_1=1$, etc. 

In particular, a uniform reference state has ${\mathcal G}^*=C_{Nv}$ and
 \begin{equation}
 T=T^1\oplus T^4\oplus T^5\oplus \cdots \oplus T^5,
 \end{equation}
 where each of the representations $T^5$ corresponds to the subspace
$L^k$, generated by the Fourier modes $({\bf e}^k)_l=\exp(\pm ikl)$ with
wavevectors $0<|k|<\pi$, $T^1$ to $k=0$ and $T^4$ to $k=\pi$. Since $T^5$
is present $M=m_5=2$. Therefore in order to control an unstable uniform
steady state of the coupled map lattice we need at least two control
parameters \cite{self}. This is the reflection of the parity symmetry of
the model (\ref{eq_cml}).

Choosing $B=({\bf b}_1\ {\bf b}_2)$ as a 2-column matrix, and defining the
Fourier coefficients
 \begin{equation}
 b^k_i={\bf e}^k\cdot {\bf b}_i,
 \end{equation}
 we write the conditions on the vectors ${\bf b}_1$ and ${\bf b}_2$:
$b^k_1\ne 0$, $b^k_2\ne 0$, $b^k_1\ne b^k_2$ for $0<|k|<\pi$ and either
$b^k_1\ne 0$ or $b^k_2\ne 0$ for $k=0,\pi$. For example, the choice
 \begin{equation}
 B_{ij}=\delta_{j,1}\delta_{i,l}+\delta_{j,2} \delta_{i,l+1}
 \end{equation}
 yields a controllable system for any $1\le l\le N$. It corresponds to
applying feedback locally through the perturbations of the variables at
the adjacent sites $l$ and $l+1$. In fact it can be easily seen that this
control arrangement makes unstable periodic orbit with arbitrary symmetry
controllable.  

All the examples above show that the symmetry is reduced upon the
linearization of the evolution equation. However the symmetry can increase
as well. It is quite easy to construct a coupled map lattice system whose
symmetry will increase for certain highly symmetric reference states. We
will see another (continuous-time) example just below.

\section{Particle in a Symmetric Potential}

The motion of a particle in a symmetric potential,
 \begin{equation}
 \label{eq_ham}
 m\partial^2_t{\bf r}=-\nabla V({\bf r}),
 \end{equation}
 serves as another example of the relation between the groups
$\mathcal{G}$ and $\mathcal{G}^*$. Suppose that the potential $V({\bf r})$
possesses the cubic symmetry (group $O\subset SO(3)$), but is not
spherically symmetric, for instance: 
 \begin{equation}
 V({\bf r})=V_0\cosh(kx)\cosh(ky)\cosh(kz).
 \end{equation}
 Linearizing the evolution equation (\ref{eq_ham}) about the equilibrium
point ${\bf r}^*=0$ we obtain 
 \begin{eqnarray}
 \label{eq_ham_lin}
 \partial_t\left(\matrix{{\bf r}\cr{\bf v}}\right)=
 \left(\matrix{0 & \openone \cr \omega^2\openone & 0}\right)
 \left(\matrix{{\bf r}\cr{\bf v}}\right),
 \end{eqnarray}
 where $\omega^2=-V_0 k^2/m$ and $\openone$ is a $3\times 3$ unit matrix. 
If $V_0<0$ the equilibrium is unstable.

Equation (\ref{eq_ham_lin}) is spherically symmetric, ${\mathcal
G}^*=SO(3)$, and therefore ${\mathcal G}\subset \mathcal{G}^*$, i.e. the
symmetry of the linearized equation is higher than the symmetry of the
original nonlinear evolution equation. 

Next we notice, that the representation $T$ of $\mathcal{G}^*$ in the
6-dimensional space $\{{\bf r},{\bf v}\}$ can be decomposed into a sum of
two 3-dimensional irreducible representations of $SO(3)$ (vector
representations): 
 \begin{equation}
 T=T^1\oplus T^1, \qquad m_1=3.
 \end{equation}
 This indicates that in order to control the unstable state ${\bf r}^*
={\bf v}^*=0$ one needs at least three independent control parameters. 

Probably the simplest way to control such a system is re-adjust the
potential (applying external fields, shifting support point, etc) based
on the instantaneous values of the position ${\bf r}$ and velocity ${\bf
v}$ of the particle. This corresponds to picking the control matrix in the
following form:
 \begin{equation}
 B=\left(\matrix{ 0 & 0 & 0 \cr {\bf b}_1 & {\bf b}_2 & {\bf b}_3}\right),
 \end{equation}
 where ${\bf b}_1,{\bf b}_2,{\bf b}_3$ could be chosen as any three
linearly independent vectors in $R^3$. 

\section{Conclusions}

Summing up, we conclude that the symmetry properties of the system should
be understood prior to constructing a control scheme. 

The number of control parameters required to control a given state of the
system can often be determined using only symmetry considerations, without
knowing anything else about the actual evolution equations. The knowledge
of the evolution equations (at least in the linearized form)  however
allows one to choose the control parameters (through the matrix $B$) 
systematically, avoiding trial-and-error search. The general idea can be
stated briefly: the controllability condition requires the control
arrangement able to break the symmetry of the evolution equation
completely. 

Care should be taken if there are accidental degeneracies. The knowledge
of the multiplicity of the degenerate eigenvalues becomes less useful and
typically leads to an overestimation of the number of control parameters
required. This case is more complicated and additional information about
the structure of the Jacobian might be necessary in order to determine the
minimal number of control parameters and construct the control matrix. 

\acknowledgements
 R.O.G is grateful to Profs. J. C. Doyle and R. Murray for valuable
discussions.  This work was partially supported by the NSF through grant
No. DMR-9013984.


\begin{references}

\bibitem{romeiras} F. J. Romeiras, C. Grebogi, E. Ott and W. P. Dayawansa,
Physica D {\bf 58}, 165 (1992). 

\bibitem{petrov} V. Petrov, E. Mihaliuk, S. K. Scott and K. Showalter,
Phys. Rev. E {\bf 51}, 3988 (1995). 

\bibitem{ding} M. Ding, W. Yang, V. In, W. L. Ditto, M. L. Spano and B.
Gluckman, Phys. Rev. E {\bf 53} 4334 (1996).

\bibitem{barreto} E. Barreto and C. Grebogi, Phys. Rev. E {\bf 52}, 3553
(1995).

\bibitem{warncke} J. Warncke, M. Bauer and W. Martienssen, Europhys. Lett.
{\bf 25}, 323 (1994).

\bibitem{self} R. O. Grigoriev, M. C. Cross and H. G. Schuster, {\sl
Pinning Control of Spatiotemporal Chaos} (submitted to Phys. Rev. Lett).

\bibitem{locher} M. Locher and E. R. Hunt, Phys. Rev. Lett. {\bf 79}, 63
(1997).

\bibitem{rubio} J. E. Rubio, {\sl The Theory of Linear Systems} (Academic 
Press, New York, 1971).

\bibitem{hammer} M. Hammermesh, {\sl Group Theory and its Application to
Physical Problems} (Addison-Wesley, Massachussets, 1964). 

\end{references}
\end{document}